\documentclass[superscriptaddress,prb,citeautoscript,longbibliography]{revtex4-1} 

\usepackage{revsymb4-1}
\usepackage{amssymb}
\usepackage{amsmath}
\usepackage{graphicx}
\usepackage{dcolumn}
\usepackage{bm} 
\usepackage{caption}
\usepackage{subcaption}
\usepackage{color}
\newcommand{\upperRomannumeral}[1]{\uppercase\expandafter{\romannumeral#1}}

\begin{document}
\title{Effect of long-range structural corrugations on magnetotransport properties of phosphorene in tilted magnetic field}

\author{A. Mogulkoc}
\email{mogulkoc@science.ankara.edu.tr}
\affiliation{Department of Physics, Faculty of Sciences, Ankara University, 06100, Tandogan, Ankara, Turkey}

\author{M. Modarresi}

\affiliation{Department of Physics, Ferdowsi University of Mashhad, Mashhad, Iran}
\affiliation{Laboratory of Organic Electronics, Department of Science and Technology, Campus Norrk\"oping, Link\"oping University, SE-60174 Norrk\"oping, Sweden}

\author{A.~N. Rudenko}
\affiliation{Radboud University, Institute for Molecules and Materials, Heyendaalseweg 135, 6525 AJ Nijmegen, The Netherlands}
\affiliation{Theoretical Physics and Applied Mathematics Department, Ural Federal University, 
Mira Str. 19, 620002 Ekaterinburg, Russia}
\date{\today}

\begin{abstract}
Rippling is an inherent quality of two-dimensional materials playing an important role in determining their properties. Here, we study the effect of structural corrugations on the electronic and transport properties of monolayer black phosphorus (phosphorene) in the presence of tilted magnetic field. 
We follow a perturbative approach to obtain analytical corrections to the spectrum of Landau levels induced by a long-wavelength corrugation potential. We show that surface corrugations have a non-negligible effect on the electronic spectrum of phosphorene in tilted magnetic field. Particularly, the Landau levels are shown to exhibit deviations from the linear field dependence. The observed effect become especially pronounced at large tilt angles and corrugation amplitudes. Magnetotransport properties
are further examined in the low temperature regime taking into account impurity scattering. We calculate magnetic field dependence of the longitudinal and Hall resistivities and find that the nonlinear effects reflecting the corrugation might be observed even in moderate fields (\mbox{$B<10$ T}).

\end{abstract}
\maketitle

\section{Introduction}
After the first synthesis of graphene \cite{Novoselov666}, the interest in two-dimensional (2D) materials has grown considerably over the past decade. The gapless energy spectrum of graphene has stimulated the search for 2D semiconductors, more suitable for traditional electronic and optoelectronic applications. Besides the other group IV materials \cite{vogt,davila2014germanene,bampoulis,zhang,zhu2015epitaxial} and a variety of transition-metal dichalcogenides\cite{wang2012electronics,deep} fabricated in recent years, new elemental materials appear in the focus of attention. In this context, few-layer black phosphorus is one of the most promising 2D materials potentially interesting for practical applications \cite{Ling2015,Gomez2015,Carvalho2016} because of its relatively high carrier mobility\cite{koenig2014electric,doi:10.1021/nn501226z,li2014black}, tunable energy gap of 0.3--2.0 eV \cite{Tran2014,Rudenko2015,Qiao2014}, and intrinsic anisotropy \cite{Qiao2014,xia2014rediscovering,Xiaomu2015} resulting in, for instance, unusual optical  response \cite{Low1,Low2}. Compared to graphene,  properties of black phosphorus are considerably less studied theoretically, which hinders the understanding of experimentally observable phenomena. 

2D materials are known to be intrinsically unstable with respect to long-wavelength thermal fluctuations, resulting in the formation of a corrugated or rippled structure in accordance with the Mermin-Wagner theorem \cite{PhysRev.176.250}. Earlier studies demonstrated that rippling is an intrinsic feature of graphene, which affects its electronic properties \cite{Meyer,Fasolino,PhysRevB.76.165409,PhysRevB.77.035423,PhysRevB.77.075422,PhysRevB.77.205421,PhysRevB.79.184205,PhysRevB.81.115421}. Other 2D structures were also shown to have a tendency to form ripples, such as, for example, in hexagonal boron nitride \cite{doi:10.1021/nl9011497}, transition metal dichalcogenides \cite{doi:10.1021/nl2022288,ADMA:ADMA201301492}, and black phosphorus \cite{doi:10.1021/acs.jpclett.5b00522,C6NR02752K,0957-4484-27-5-055701}. Although 2D materials are usually deposited on substrates, which may suppress the formation of intrinsic rippling, surface roughness of common dielectrics like SiO$_2$ represents by itself another source of structural corrugations \cite{Ishigami,Geringer,Heinz}.

Understanding the dynamics of charge carriers in 2D materials under realistic conditions is a problem of practical importance as it determines observable transport properties. Magnetotransport measurements offer a powerful tool to probe carrier dynamics at the quantum level. Recently, several studies have reported quantum transport measurements in few-layer black phosphorus \cite{chen2015,li2015quantum,tayari2015,gillgren2015,li2016quantum,long2016,tran2017surface,long2017}. Interpretation of experimental observations is usually carried out on a phenomenological level without explicit consideration of their microscopic nature. On the other hand, theoretical description of quantum transport at the level of model Hamiltonians\cite{yuan2015,zhou2015landau,PhysRevB.92.045420,PhysRevB.92.085408,PhysRevB.92.165405,yuan2016} has limited capability to capture essential environmental effects caused by impurities, substrates, and structural corrugations. The role of those effects in magnetotransport properties of few-layer black phosphorus is not well understood.

In this paper, we study the role of long-range structural corrugations on the Landau levels (LLs) and magnetotransport properties of monolayer black phosphorus (MBP) in the presence of a tilted magnetic field.  We use a perturbative approach to obtain first-order corrections to the energy spectrum induced by a corrugation potential. We find noticeable deviations of LLs from the linear dependence on magnetic field, which are also apparent in the calculated longitudinal and Hall resistivities at not very strong fields.

The paper is organized as follows. The theory part is presented in Sec.~II, where we first consider unperturbed Hamiltonian for MBP in perpendicular magnetic field (Sec.~II~A), and then obtain a correction to the Hamiltonian in the presence of a corrugation potential in tilted magnetic field (Sec.~II~B).  In Sec.~II~C, we present the formalism of the linear response theory, which is used to calculate magnetotransport properties of MBP. The results and their discussion are presented in Sec.~III. In Sec.~IV, we briefly summarize our results and conclude the paper.

\section{Theory}
\subsection{Pristine MBP in perpendicular magnetic field}
The energy spectrum of MBP can be described by a four-band tight-binding model \cite{PhysRevB.89.201408}. However, $C_{2h}$ group invariance of the MBP lattice enables one to describe the system by a two-band model \cite{PhysRevLett.112.176801}. An effective continuum model can be obtained by expanding the tight-binding Hamiltonian around the $\mathrm{\Gamma} $ point, yielding a good agreement with the tight-binding results in the energy range of $\sim$3.5 eV \cite{1367-2630-16-11-115004,PhysRevB.92.075437}. In the long-wavelength limit, the continuum Hamiltonian of MBP can be written as \cite{PhysRevB.92.075437}, 

\begin{equation}
\mathcal{H}_{0}=\left[ 
\begin{array}{cc}
u_{0}+\bar{\eta}_{x}\pi_{x}^{2}+\bar{\eta}_{y}\pi_{y}^{2} & \delta+\bar{\gamma}_{x}\pi_{x}^{2}+\bar{\gamma}_{y}\pi_{y}^{2}+i\bar{\chi}\pi_{y} \\ 
\delta+\bar{\gamma}_{x}\pi_{x}^{2}+\bar{\gamma}_{y}\pi_{y}^{2}-i\bar{\chi}\pi_{y} & u_{0}+\bar{\eta}_{x}\pi_{x}^{2}+\bar{\eta}_{y}\pi_{y}^{2}
\end{array}%
\right].  \label{1}
\end{equation}%
Eq.~(\ref{1}) can be divided into quadratic and linear terms, respectively, as
\begin{equation}
	\mathcal{H}_{0}^{\mathrm{quad}}=\left[ 
	\begin{array}{cc}
		u_{0}+\bar{\eta}_{x}\pi_{x}^{2}+\bar{\eta}_{y}\pi_{y}^{2} & \bar{\gamma}_{x}\pi_{x}^{2}+\bar{\gamma}_{y}\pi_{y}^{2} \\ 
		\bar{\gamma}_{x}\pi_{x}^{2}+\bar{\gamma}_{y}\pi_{y}^{2} & u_{0}+\bar{\eta}_{x}\pi_{x}^{2}+\bar{\eta}_{y}\pi_{y}^{2}
	\end{array}%
	\right],  \label{2}
\end{equation}%

\begin{equation}
	\mathcal{H}_{0}^{\mathrm{lin}}=\left[ 
	\begin{array}{cc}
		0 & \delta+i\bar{\chi}\pi_{y} \\ 
		\delta-i\bar{\chi}\pi_{y} & 0
	\end{array}%
	\right].  \label{3}
\end{equation}%
 Here, $ \bar{\eta}_i=\eta_{i}/\hbar^{2} $ ($ \eta_{x}=0.58 $ eV\AA{}$ ^{2} $ and $ \eta_{y}=1.01 $ eV\AA{}$ ^{2} $), $ \bar{\gamma}_i=\gamma_{i}/\hbar^{2} $ ($ \gamma_{x}=3.93 $ eV\AA{}$ ^{2} $ and $ \gamma_{y}=3.83 $ eV\AA{}$ ^{2} $), $ \bar{\chi}=\chi/\hbar $ ($ \chi=5.25 $ eV\AA{}), $ u_{0}=-0.42 $ eV and $ \delta=0.76 $ eV \cite{PhysRevB.92.075437}, and $\pi_{i}$ is the 2D canonical momentum. If magnetic field is applied normal to the MBP plane, {\bf B}=(0,0,B), in symmetric gauge $\pi_{x}=p_{x}-(eB/2)y$ and $\pi_{y}=p_{y}+(eB/2)x$, where $p_{i}$ is the momentum operator. One can express $p_{i}$ and $r_{i}$ in terms of the creation $b_i^{\dag}$ and annihilation $b_i$ operators as
\begin{eqnarray}
p_{i}&=&\left(\frac{m^{\lambda}_{i} \hbar \omega_{\lambda}}{2}\right)^{1/2}(b_{i}^{\dagger}+b_{i}), \notag \\
r_{i}&=&-i\left(\frac{\hbar }{2m^{\lambda}_{i} \omega_{\lambda}}\right)^{1/2}(b_{i}^{\dagger}-b_{i}), \notag
\end{eqnarray}%
where $\lambda$ is the band index taking $+$1 ($-$1) for the conduction (valence) band, and $i$ refers to $x$ or $y$. $\omega_{\lambda}=eB/\sqrt{m^{\lambda}_{x}m^{\lambda}_{y}}$ is the cyclotron frequency, which takes $\omega_{+}=2.668 \omega_{e}$ ($\omega_{-}=2.195 \omega_{e}$) for electrons (holes) with $\omega_{e}=eB/m_0$, and $m^{\lambda}_i$ are the effective masses: $m^{+}_{x}=\hbar^{2}/2(\eta_{x}+\gamma_{x})=0.846 m_0$ and $m^{+}_{y}=\hbar^{2}/2(\eta_{y}+\gamma_{y}+\chi^{2}/2\delta)=0.166 m_0$ for the conduction band, $m^{-}_{x}=\hbar^{2}/2(\gamma_{x}-\eta_{x})=1.140 m_0$ and $m^{-}_{y}=\hbar^{2}/2(\eta_{y}-\gamma_{y}-\chi^{2}/2\delta)=0.182 m_0$ for the valance band, with $m_0$ being the free electron mass. By diagonalizing $\mathcal{H}_{0}^{\mathrm{quad}}$, eigenvalues of the quadratic Hamiltonian become
\begin{eqnarray}
E_{n}^{\mathrm{quad}}&=&u_{0}+(\bar{\eta}_x+\lambda \bar{\gamma}_x)\left\langle n_{x}n_{y} \right| \pi_{x}^{2} \left| n_{x}n_{y}\right\rangle+(\bar{\eta}_y+\lambda \bar{\gamma}_y)\left\langle n_{x}n_{y} \right| \pi_{y}^{2} \left| n_{x}n_{y}\right\rangle,
\label{4}
\end{eqnarray}%
where
\begin{eqnarray}
\left\langle n_{x}n_{y} \right| \pi_{x}^{2} \left| n_{x}n_{y}\right\rangle&=&\frac{m^{\lambda}_{x} \hbar \omega_{\lambda}}{2}(b_{x}^{\dagger}+b_{x})^{2}-\left(\frac{eB}{2}\right)^{2}\frac{\hbar}{2m^{\lambda}_{y}\omega_{\lambda}}(b_{y}^{\dagger}-b_{y})^{2} \notag \\
&-&ieB \sqrt{\frac{m^{\lambda}_{x} \hbar \omega_{\lambda}}{2}}\sqrt{\frac{\hbar}{2m^{\lambda}_{y}\omega_{\lambda}}}(b_{x}^{\dagger}+b_{x}) (b_{y}^{\dagger}-b_{y}), \notag \\
\left\langle n_{x}n_{y} \right| \pi_{y}^{2} \left| n_{x}n_{y}\right\rangle&=&\frac{m^{\lambda}_{y} \hbar \omega_{\lambda}}{2}(b_{y}^{\dagger}+b_{y})^{2}-\left(\frac{eB}{2}\right)^{2}\frac{\hbar}{2m^{\lambda}_{x}\omega_{\lambda}}(b_{x}^{\dagger}-b_{x})^{2} \notag \\
&+&ieB \sqrt{\frac{m^{\lambda}_{y} \hbar \omega_{\lambda}}{2}}\sqrt{\frac{\hbar}{2m^{\lambda}_{x}\omega_{\lambda}}}(b_{y}^{\dagger}+b_{y}) (b_{x}^{\dagger}-b_{x}).
\end{eqnarray}%
In turn, diagonalization of the linear term $\mathcal{H}_{0}^{\mathrm{lin}}$ yields
\begin{eqnarray}
(E_{n}^{\mathrm{lin}})^{2}&=&\delta^{2}+ \bar{\chi}^{2}\left\langle n_{x}n_{y} \right| \pi_{y}^{2} \left| n_{x}n_{y}\right\rangle.
\label{5}
\end{eqnarray}%
Due to the gauge independent degeneracy of LLs, we assume $n_x=n_y=n$, thus eigenvalues of the total Hamiltonian $\mathcal{H}_{0}$ can be written as (see Appendix A for more details)
\begin{eqnarray}
E_{n\lambda}^{0}&=&E_{n}^{\mathrm{quad}}+E_{n}^{\mathrm{lin}}\notag \\
&=&u_{0}+\lambda \left[ \left| (\bar{\eta}_x+\lambda \bar{\gamma}_x)\right|m^{\lambda}_{x} +\left| (\bar{\eta}_y+\lambda \bar{\gamma}_y)\right|m^{\lambda}_{y} \right] \hbar \omega_{\lambda} \left(n+\frac{1}{2}\right)+\lambda \left[\delta^{2}+ \bar{\chi}^{2} m^{\lambda}_{y} \hbar \omega_{\lambda} \left(n+\frac{1}{2}\right)\right]^{\frac{1}{2}}. 
\label{6}
\end{eqnarray}%
The last term can be expanded as $\delta \left[1+ (\bar{\chi}^{2} m^{\lambda}_{y}/\delta^{2}) \hbar \omega_{\lambda} (n+1/2)\right]^{1/2}\approx \left[\delta+ (\bar{\chi}^{2} m^{\lambda}_{y}/2\delta) \hbar \omega_{\lambda} (n+1/2)\right]$. Finally, we arrive at
\begin{eqnarray}
E_{n\lambda}^{0}&=&(u_{0}+\lambda \delta)+\lambda \hbar \omega_{\lambda} \left(n+\frac{1}{2}\right). 
\label{7}
\end{eqnarray}%
\begin{figure} [h]
    \centering
    \includegraphics[width=0.7\textwidth]{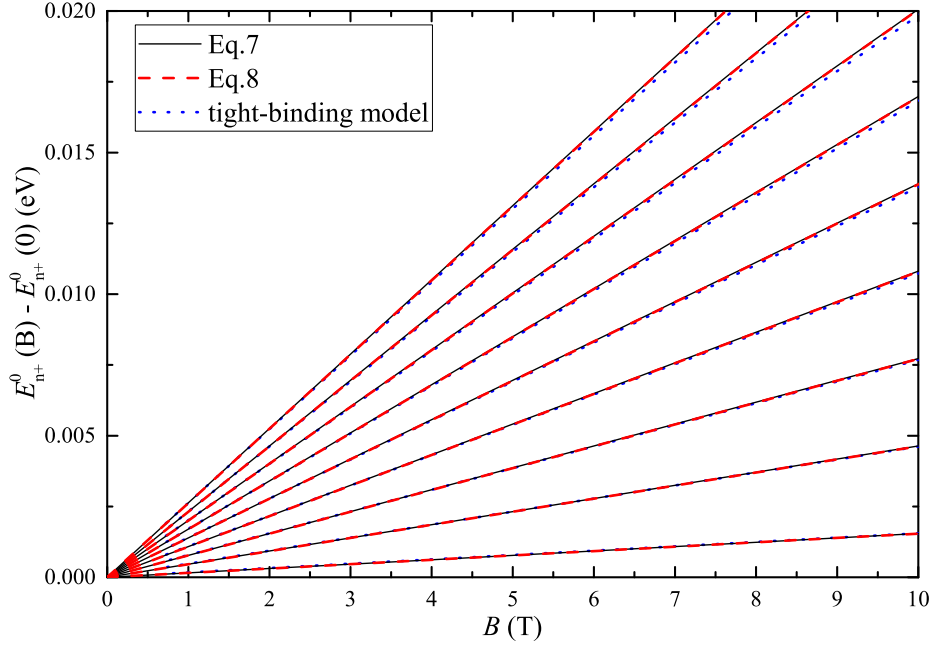}
    \caption{Landau quantization of electron states in MBP. Black line corresponds to Eq.~(\ref{6}), red dashed line corresponds to Eq.~(\ref{7}). Blue dotted line corresponds to the calculations within a tight-binding model (see Appendix B for details).}
    \label{FIGURE1}
\end{figure}
The expression given by Eq.~(\ref{7}) is fully consistent with the results of previous studies \cite{PhysRevB.92.075437,zhou2015landau}. It is clear from Fig.~\ref{FIGURE1} that for $B<10$ T, two expressions in Eqs.~(\ref{6}) and (\ref{7}) match with each other demonstrating that the linear term ($\chi$) in the continuum Hamiltonian is less effective on LLs of MBP.  From Fig.~\ref{FIGURE1} one can also see that both spectra are very close to the results of tight-binding calculations performed in Appendix B. In Appendix B, we also consider the case of in-plane magnetic field, which is shown to have a negligible effect on the properties of pristine (noncorrugated) MBP.

\subsection{Corrugated MBP in tilted magnetic field}
We now consider a vector potential that produces a tilted magnetic field,
\begin{equation}
\mathbf{B}=\left[\frac{B_{\parallel}+B_{\perp}\sin\theta}{\sqrt{2}},\frac{B_{\parallel}+B_{\perp}\sin\theta}{\sqrt{2}},B_{\perp}\cos\theta\right], 
\end{equation}
which consists of a constant field $B_{\parallel}$ along the $xy$-plane and a constant field $B_{\perp}$ tilted with respect to the $z$-axis by angle $\theta$.  Modified symmetric gauge which yields this magnetic field can be chosen as
\begin{eqnarray}
\mathbf{A}&=&\left[\frac{-yB_{\perp}\cos \theta}{2}+\frac{z(B_{\parallel}+B_{\perp}\sin \theta)}{\sqrt{2}},\frac{xB_{\perp}\cos \theta}{2}-\frac{z(B_{\parallel}+B_{\perp}\sin \theta)}{\sqrt{2}},0 \right].
\label{A}
\end{eqnarray}%
Similar gauge choices were considered before for parabolic quantum wells \cite{0268-1242-9-5S-110,HAUPT1994214,PhysRevB.60.8984} and other 2D materials \cite{Kandemir2010,PSSB:PSSB201552341}.
In Eq.~(\ref{A}), even if the tilt angle $\theta$ is set to zero, the parallel component $B_{\parallel}$ still exists which, allows us to examine the effect of $B_{\parallel}$ on the energy spectrum of MBP. In the presence of tilted magnetic field, the square of the momentum operators is given by
\begin{eqnarray}
\pi_{x}^{2}&=&\left(p_{x}-\frac{eB_{\perp}\cos \theta}{2} y\right)^{2}+\frac{e^{2}z^{2}(x,y)}{2}\Theta^{2}(B,\theta)+\left(p_{x}-\frac{eB_{\perp}\cos \theta}{2} y\right)\left(\frac{ez(x,y)}{\sqrt{2}}\Theta(B,\theta)\right) \notag \\
&+&\left(\frac{ez(x,y)}{\sqrt{2}}\Theta(B,\theta)\right)\left(p_{x}-\frac{eB_{\perp}\cos \theta}{2} y\right) \notag \\ 
\pi_{y}^{2}&=&\left(p_{y}+\frac{eB_{\perp}\cos \theta}{2} x\right)^{2}+\frac{e^{2}z^{2}(x,y)}{2}\Theta^{2}(B,\theta)-\left(p_{y}+\frac{eB_{\perp}\cos \theta}{2} x\right)\left(\frac{ez(x,y)}{\sqrt{2}}\Theta(B,\theta)\right) \notag \\
&-&\left(\frac{ez(x,y)}{\sqrt{2}}\Theta(B,\theta)\right)\left(p_{y}+\frac{eB_{\perp}\cos \theta}{2} x\right).
\label{Picorrug}
\end{eqnarray}%
\begin{figure} [h]
    \centering
    \includegraphics[width=0.6\textwidth]{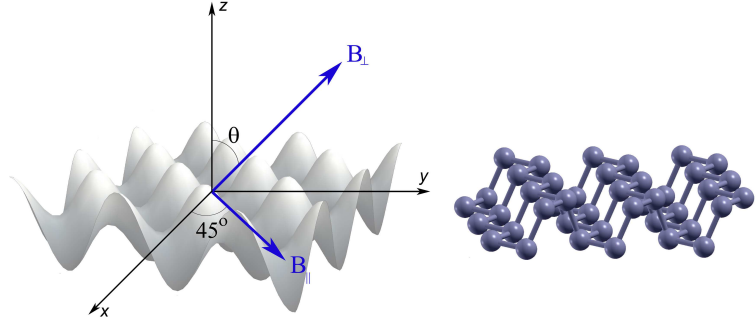}
    \caption{Left: Schematic representation of a corrugation potential in the presence of a tilted perpendicular ($B_{\perp}$) and in-plane ($B_{\parallel}$) magnetic fields. Right: Puckered structure of MBP.}
    \label{FIGURE2}
\end{figure}
In Eq.~(\ref{Picorrug}), we have introduced a corrugation potential along the $xy$-plane having the form $z(x,y)=V\cos(\textrm{K}x)\cos(\textrm{K}^{\prime}y)$ which can be considered as a small perturbation on the surface of MBP (see Fig.\ref{FIGURE2}).
Here, $\textrm{K}=2\pi/\ell_{x}$ and $\textrm{K}^{\prime}=2\pi/\ell_{y}$, $\ell_{x}$ and $\ell_{y}$ are the length of the corrugation along the $x$ and $y$ directions, respectively. $V$ is the height (amplitude) of the corrugation, $\Theta(B,\theta)=B_{\perp}(\sin \theta+\xi)$, and $\xi=B_{\parallel}/B_{\perp}$. 
 In what follows, the effect of corrugation potential on LLs is treated perturbatively, and assuming $B_{\perp}>B_{\parallel}$ ($\xi<1$), which preserves $C_{2h}$ group invariance of the MBP lattice. 
More sophisticated analysis can be, in principle, performed following the variational technique \cite{kandemir2010variational,kandemir2010boundaries}.
Considering the modified momentum operators in Eq.~(\ref{Picorrug}) and following the same procedure outlined in Sec.~II~A, the energy eigenvalues of the system can be written as
\begin{eqnarray}
E_{n\lambda}&=&\overline{E}_{n\lambda}^{0}+\frac{\lambda}{2} \Delta E_{n},
\label{8}\\
\Delta E_{n}&=&\left(\Delta E_{n}^{x}+\Delta E_{n}^{y}\right) \notag
\end{eqnarray}%
Here, $\overline{E}_{n\lambda}^{0}$ is a modified angle-dependent version of the energy eigenvalues that appeared in Eq.~(\ref{7}), i.e., $\overline{E}_{n\lambda}^{0}=(u_{0}+\lambda \delta)+\lambda \hbar \overline{\omega}_{\lambda} (n+1/2)$, where $\overline{\omega}_{\lambda}=\omega_{\lambda}\cos \theta$ is the modified cyclotron frequency. $\Delta E_{n}^{x}$ and $\Delta E_{n}^{y}$ are the first-order corrections to the energy eigenvalues given by
\begin{eqnarray}
\Delta E_{n}^{i}&=&\frac{e^{2}B_{\perp}^{2}}{2m^{\lambda}_{i}}\left(\sin \theta +\xi\right)^{2}\frac{V^{2}}{4}\mathbb{G}_{n},
\label{DeltaE}
\end{eqnarray}
with
\begin{eqnarray}
\mathbb{G}_{n}&=&\left\langle n_{x}n_{y} \right| \cos^{2}(\textrm{K}x)\cos^{2}(\textrm{K}^{\prime}y) \left| n_{x}n_{y}\right\rangle
\end{eqnarray}%
being the spatial correlation function defined as
\begin{eqnarray}
\mathbb{G}_{n}&=&\int_{-\infty}^{\infty} \int_{-\infty}^{\infty} dx dy \Psi_{n_{x}n_{y}}^{*}(x,y) \cos^{2}(\textrm{K}x)\cos^{2}(\textrm{K}^{\prime}y) \Psi_{n_{x}n_{y}}(x,y).
\label{9}
\end{eqnarray}%
In Eq.~(\ref{9}), $\Psi_{n_{x}n_{y}}(x,y) =\left\langle \mathbf{r} |n_{x}n_{y}\right\rangle$, and
\begin{eqnarray}
\left\langle \mathbf{r} |n_{x}n_{y}\right\rangle&=&\frac{1}{\sqrt{2^{n_{x}}n_{x}!\sqrt{\pi}}}\frac{1}{\sqrt{2^{n_{y}}n_{y}!\sqrt{\pi}}}\sqrt[4]{\frac{m^{\lambda}_{x} \overline{\omega}_{\lambda}} {\hbar}}\sqrt[4]{\frac{m^{\lambda}_{y} \overline{\omega}_{\lambda}} {\hbar}}\notag \\
&\times& \exp\left(\frac{-m^{\lambda}_{x} \overline{\omega}_{\lambda}} {2\hbar}x^{2}\right) \exp\left(\frac{-m^{\lambda}_{y} \overline{\omega}_{\lambda}} {2\hbar}y^{2}\right) \notag \\
&\times&H_{n_{x}}\left(\sqrt{\frac{m^{\lambda}_{x} \overline{\omega}_{\lambda}} {\hbar}}x\right)H_{n_{y}}\left(\sqrt{\frac{m^{\lambda}_{y} \overline{\omega}_{\lambda}} {\hbar}}y\right),
\label{wf}
\end{eqnarray}%
where $H_{n}$ are the Hermite polynomials (see Appendix A). After taking the integrals in Eq.~(\ref{9}) and considering the assumption $n_x=n_y=n$, we get (see Appendix for more details)
\begin{eqnarray}
\mathbb{G}_{n}&=&\left[1+\exp\left(\frac{-\textrm{K}^{2}\hbar}{m^{\lambda} _{x}\overline{\omega}_{\lambda}}\right)L_{n}\left(\frac{2\textrm{K}^{2}\hbar}{m^{\lambda}_{x}\overline{\omega}_{\lambda}}\right)\right]\left[1+\exp\left(\frac{-\textrm{K}^{\prime 2}\hbar}{m^{\lambda}_{y}\overline{\omega}_{\lambda}}\right)L_{n}\left(\frac{2\textrm{K}^{\prime 2}\hbar}{m^{\lambda}_{y}\overline{\omega}_{\lambda}}\right)\right],
\label{10}
\end{eqnarray}%
where $L_{n}$ are the Laguerre polynomials. To see the oscillatory nature of the Laguerre polynomials, their asymptotic expression can be used, $e^{u/2}L_{n}(u)\approx \left(\pi^{2}nu\right)^{-1/4}\cos \left(2\sqrt{nu}-\pi/4\right)$. For large $n$, $n\rightarrow(E_{F}/\hbar \omega_{\lambda})-1/2$ \cite{PhysRevLett.63.2120,PhysRevB.47.1466} and
\begin{eqnarray}
\mathbb{G}_{n}&\approx&\left[1+\frac{1}{\sqrt{\pi}}\frac{1}{[(\pi \hbar n_{c}/m_{\lambda } \overline{\omega}_{\lambda})-1/2]}
\left(\frac{m^{\lambda}_{x}\overline{\omega}_{\lambda}}{2\textrm{K}^{2}\hbar}\right)^{1/4}\cos \left[2\left[\left(\frac{\pi \hbar n_{c}}{m_{\lambda } \overline{\omega}_{\lambda}}-\frac{1}{2}\right)\frac{m^{\lambda}_{x}\overline{\omega}_{\lambda}}{2\textrm{K}^{2}\hbar}\right]^{1/2}-\frac{\pi}{4}\right]\right] \notag \\
&\times&\left[1+\frac{1}{\sqrt{\pi}}\frac{1}{[(\pi \hbar n_{c}/m_{\lambda } \overline{\omega}_{\lambda})-1/2]}
\left(\frac{m^{\lambda}_{y}\overline{\omega}_{\lambda}}{2\textrm{K}^{\prime 2}\hbar}\right)^{1/4}\cos \left[2\left[\left(\frac{\pi \hbar n_{c}}{m_{\lambda } \overline{\omega}_{\lambda}}-\frac{1}{2}\right)\frac{m^{\lambda}_{y}\overline{\omega}_{\lambda}}{2\textrm{K}^{\prime 2}\hbar}\right]^{1/2}-\frac{\pi}{4}\right]\right].
\end{eqnarray}%
Here, $m_{\lambda}=(m_x^{\lambda}m_y^{\lambda})^{1/2}$ is the cyclotron mass, and $n_{c}=m_{\lambda} E_{F}/\pi \hbar^{2}$ is the carrier concentration.

The density of states (DOS) for quantized energy spectrum can be calculated as
\begin{eqnarray}
D(E)&=&\frac{1}{S}\sum_{n,\lambda}\delta \left(E-E_{n \lambda}\right) 
\label{11}
\end{eqnarray}%
where $S$ is the area of the MBP unit cell. To calculate DOS, we use the Gaussian functions as an approximation to the Dirac function in Eq.~(\ref{11}), i.e., $\delta \left(E-E_{n \lambda}\right)\approx \left(1/\sigma \sqrt{\pi}\right) \exp\left[-\left(E-E_{n \lambda}\right)^{2}/\sigma^{2}\right]$, with $\sigma$ being the broadening parameter taken to be $\sigma=0.1$ meV.

\subsection{Magnetotransport properties}

To examine the effect of tilted magnetic field on magnetotransport properties of MBP, we make use of the linear response theory. We consider a strongly quantized regime, in which $\overline{\omega}_{\lambda}\gg \tau^{-1}$, where $\tau$ is the carrier relaxation time. In the presence of the perturbative term in Eq.~(\ref{8}), the carrier velocity along both $x$ and $y$ directions remains zero due to the Landau quantization.
In this situation, one can distinguish between the two main contributions to the conductivity tensor, namely, transverse (Hall) $\sigma_{xy}$ and longitudinal (collisional) conductivity $\sigma_{xx}$. The Hall conductivity can be readily evaluated as
\begin{eqnarray}
\sigma_{xy}&=&g_{s}\frac{e^{2}}{h}\sum_{n=0}^{\infty}\sum_{\lambda=\pm}\left(n+1\right)\left[f\left(E_{n, \lambda}\right)-f\left(E_{n+1, \lambda}\right)\right],
\label{12}
\end{eqnarray}%
which is a standard expression for conventional 2D electron gas \cite{PhysRevB.32.771,PhysRevLett.63.2120,PhysRevB.46.4667,PhysRevB.47.1466}. Here, $g_{s}=2$ stands for the spin degrees of freedom, and $f\left(E_{n, \lambda}\right)=\left[1+\exp \beta \left(E_{n, \lambda}-E_{F}\right) \right]^{-1}$ is the Fermi-Dirac distribution function, where $E_{F}$ is the Fermi energy, $\beta=1/k_{B}T$ is the inverse temperature in energy units with $k_{B}$ being the Boltzmann constant. It is worth noting that $\sigma_{xy}$ is scattering independent.

The second contribution to the conductivity tensor, i.e., longitudinal conductivity, can be evaluated as \cite{PhysRevB.32.771,PhysRevLett.63.2120,PhysRevB.46.4667,PhysRevB.47.1466,PhysRevB.34.1057}
\begin{eqnarray}
\sigma_{xx}&=&g_{s}\frac{e^{2}}{h}\frac{n^{1/2}_{\mathrm{imp}}\beta}{4\pi^{3/2} \overline{\ell}_{B}}\sum_{n=0}^{\infty}\sum_{\lambda=\pm}U^{\lambda}(2n+1)f\left(E_{n, \lambda}\right) \left[1-f\left(E_{n, \lambda}\right)\right]. 
\label{13}
\end{eqnarray}%
Here, we assume scattering on randomly distributed Coulomb impurities with density $n_{\mathrm{imp}}$. Other scattering mechanism, such as scattering on phonons can be neglected in the limit of low temperatures. In Eq.~(\ref{13}), $\overline{\ell}_{B}=\sqrt{\hbar/eB_{\perp}\cos \theta}$ is the magnetic length, and $U^{\lambda}=2\pi e^{2} k_e/\epsilon k_{s}^{\lambda}$ is the impurity Coulomb potential for small momentum transfer $q\ll k_s$, where $\epsilon$ is the relative dielectric permittivity, $k_e=1/4\pi\epsilon_0$ is the Coulomb constant, and $k_{s}^{\lambda}=2\pi e^2 D^{\lambda}_0$ is the screening wave vector of electrons (holes) in the Thomas-Fermi approximation \cite{Katsnelson-Book} with $D^{\lambda}_0=m_{\lambda}/\pi\hbar^2$ being DOS in the absence of magnetic field. Eq.~(\ref{13}) represents a collisional contribution to the conductivity $\sigma_{xx}=\sigma_{xx}^{\mathrm{col}}$, which increases with impurity concentration $n_{\mathrm{imp}}$ contrary to the diffusive contribution $\sigma_{xx}^{\mathrm{dif}} \sim 1/n_{\mathrm{imp}}$ \cite{yuan2015,PhysRevB.93.165402}. This is because impurity scattering in the presence of a magnetic field favors electron hoppings between quantized cyclotron orbits \cite{PhysRevB.46.4667}, thus increasing the conductivity. It is also interesting to note that as long as $\sigma_{xx}^{\mathrm{col}}$ dominates, $\sigma_{xx}$ remains isotropic. We note that in contrast to earlier studies \cite{zhou2015landau}, we explicitly take into account impurity- and field-induced broadening of the LLs width\cite{PhysRevB.34.1057}. Using Eq.~(\ref{13}), the Hall and longitudinal resistivity can be calculated as $\rho_{xy}=\sigma_{xy}/S$ and $\rho_{xx}=\sigma_{xx}/S$, respectively, where $S=\sigma_{xx}\sigma_{yy}-\sigma_{xy}\sigma_{yx}\approx\sigma_{xy}^{2}$ assuming $\sigma_{xy}\gg\sigma_{xx}$ in sufficiently strong fields. In the following magnetotransport calculations, we use $T=1$ K, $n_{\mathrm{imp}}=10^{12}$ cm$^{-2}$, and $\epsilon=1$. The latter corresponds to the case of freestanding non-doped MBP \cite{Prishchenko}.

\section{Results and Discussion}

\begin{figure} [h]
    \centering
    \includegraphics[width=\textwidth]{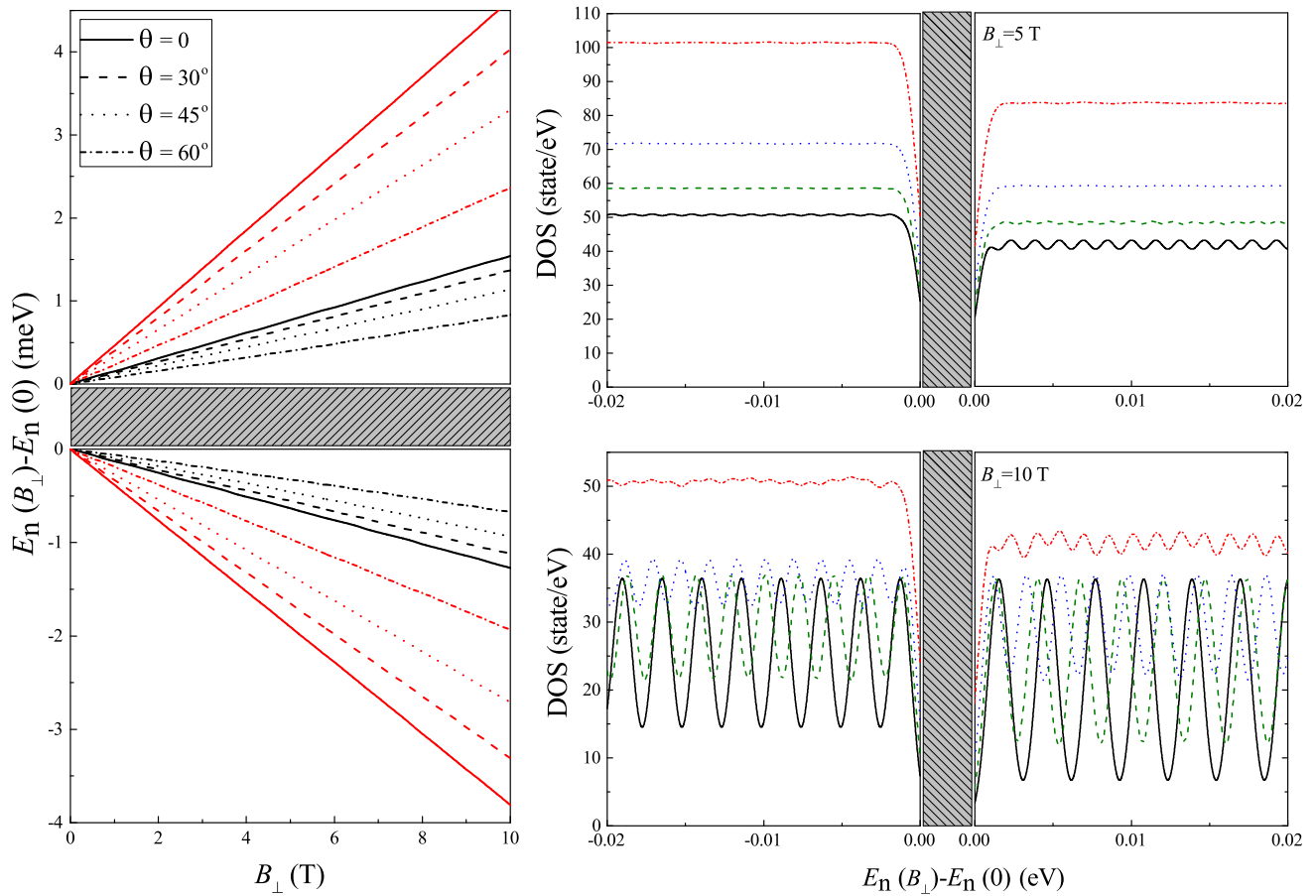}
    \caption{Left panel: Field-dependence of the two first LLs in MBP calculated for different tilt angles $\theta$. Black line corresponds to $n=0$, red line to $n=1$. Right panel: Density of states (DOS) in the vicinity of a gap (shaded area) calculated for $B_{\perp}=5$ T (top) and $B_{\perp}=10$ T (bottom) for different $\theta$ and fixed $\xi= B_{\parallel}/B_{\perp}=0.5$. All cases with $\theta \neq 0$ correspond to the corrugation potential with amplitude $V=1$ \AA {} and lengths $l_1=l_2=250$ \AA.}
    \label{FIGURE3}
\end{figure}

We first examine evolution of LLs in MBP considering tilted magnetic field in the presence of a corrugation potential. 
Since we use a perturbative approach to describe the effect of angle-dependent magnetic field [Eq.~(\ref{8})], the product $(B_{\perp}V)^2$ must not be too large to ensure validity of the approach, that is to satisfy $\Delta E_{n\lambda} \ll E_n$. This condition holds for $B_{\perp}<10$ T and $V<5$ \AA{} considered in this work.
Unless stated otherwise, we consider fixed ratio $\xi=B_{\parallel}/B_{\perp}=0.5$, and the corrugation length along both directions $\ell_{1}=\ell_{2}=250$ \AA{}, which is an order of intrinsic ripples length in graphene \cite{Meyer,Fasolino}. The case $\theta=0$ is evaluated for $V=0$ to be consistent with the results of earlier works \cite{PhysRevB.92.075437,zhou2015landau}. 

In the left panel of Fig.~\ref{FIGURE3}, we show the Landau level diagram calculated for both electron and hole states for different values of tilt angle $\theta$, and fixed amplitude of the corrugation potential $V=1$ \AA. One can see that energies of LLs decrease with $\theta$, while the linearity of the curves is preserved in the regime of relatively small corrugations and not too strong magnetic fields. 
The effect of the tilt angle on LLs is twofold. While $B_{\perp}$ confines the motion of charge carriers in the $xy$ plane, changing the magnetic field direction increases the cyclotron radius in the $xy$ plane due to the $\mathrm{cos}\theta$ factor in $\overline{\omega}_{\lambda}$. As a result, the LL energies $E_n$ decrease with $\theta$, which effectively correspond to a smaller magnetic field. This effect is partially compensated by the presence of the corrugation potential, which provides an additional contribution $\Delta E_n$ to $E_n$ [Eq.~(\ref{DeltaE})].
As can be inferred from Fig.~\ref{FIGURE3}, the main contribution to LLs comes from the first term, $\overline{E}_{n\lambda}^{0}$ in Eq.~(\ref{8}), which is strongly dependent on the perpendicular component of the out-of-plane magnetic field $B_{\perp} \cos \theta$. The effect of the second term, $ \Delta E_{n}$ in Eq.~(\ref{8}) is small for corrugations as low as $V=1$ \AA. The effect of tilted magnetic field on the electronic spectrum can be seen also from DOS shown for different tilt angles at $B_{\perp}=5$ T and $B_{\perp}=10$ T (right panel of Fig.~\ref{FIGURE3}). At $\theta \neq 0$, 
the energy spacing between LLs becomes smaller, which leads to more dense electronic states and larger DOS. A similar effect of tilted magnetic field on LLs of graphene was reported previously\cite{Kandemir2010,doi:10.1063/1.4818605}. For larger values of $B_{\perp}$, the energy spacing between LLs decreases, which gives rise to more pronounced oscillations in DOS shown in Fig.~\ref{FIGURE3} for $B_{\perp}=10$ T.

\begin{figure} [h!]
	\centering
	\includegraphics[width=0.9\textwidth]{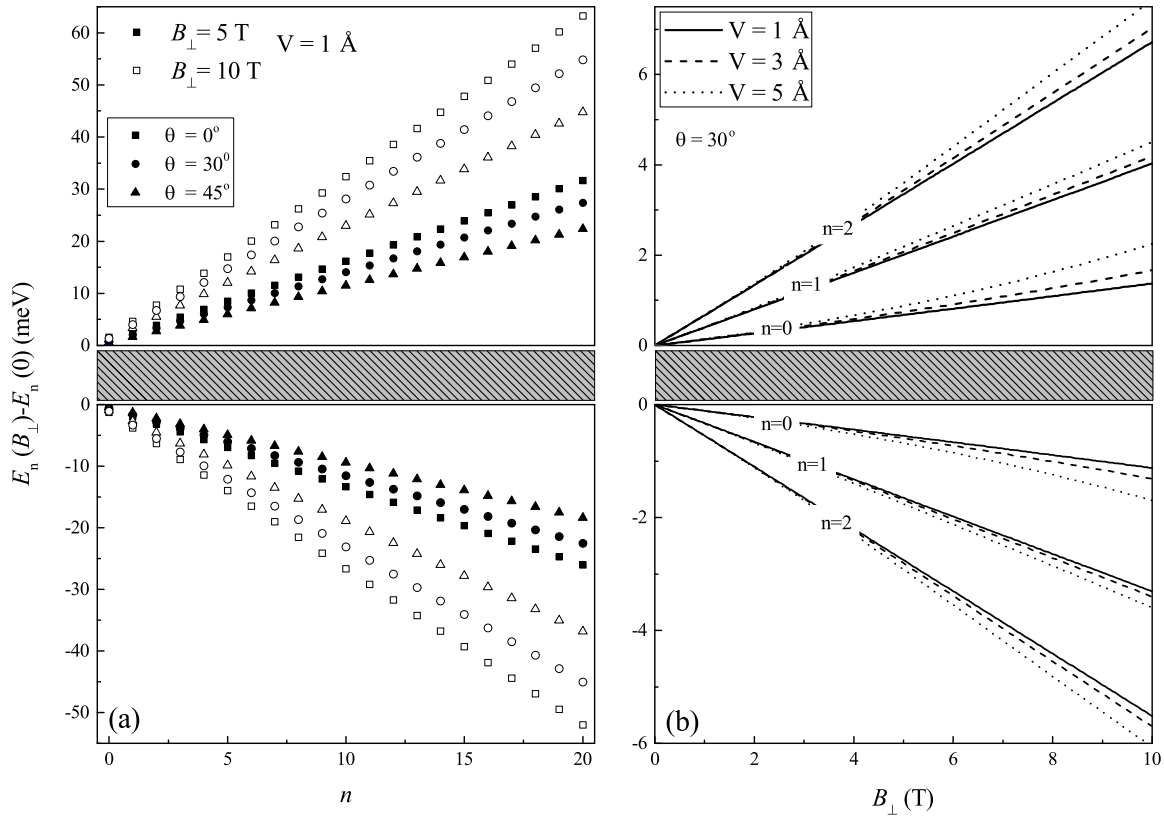}
		\caption{Energies of LLs shown as a function of the level index $n$ for different tilt angles $\theta$, magnetic fields $B_{\perp}$ and corrugation heights $V$. Shaded area is an energy gap.}
	\label{FIGURE4}
\end{figure}

In Fig.~\ref{FIGURE4}(a), the index ($n$) dependence of LLs is shown in the presence of tilted magnetic field for $V=1$ \AA{}. LLs splitting of electron and hole states is different because of the electron-hole asymmetry and unequal effective masses. 
In Fig.~\ref{FIGURE4}(b), the magnetic field ($B_{\perp}$) dependence of LLs is shown for different corrugation heights at $\theta=30$ \AA{}. One can see pronounced deviations from the linear behavior, which become especially clear for $V=5$ \AA{} and $B_{\perp}=10$ T. In the chosen range of parameters, these deviations do not exceed $\hbar \overline{\omega}_{\lambda}(n+1/2)$, demonstrating the validity of the perturbative approach. The observed nonlinearity is a manifestation of long-range structural corrugations. Although at relatively weak fields, first-order correction to the LLs energy is quadratic in $V$ [Eq.~(\ref{DeltaE})], the dependence at large fields may be different. Particularly, we do not exclude oscillatory behavior in this regime. 

\begin{figure} [h!]
	\centering
	\includegraphics[width=0.8\textwidth]{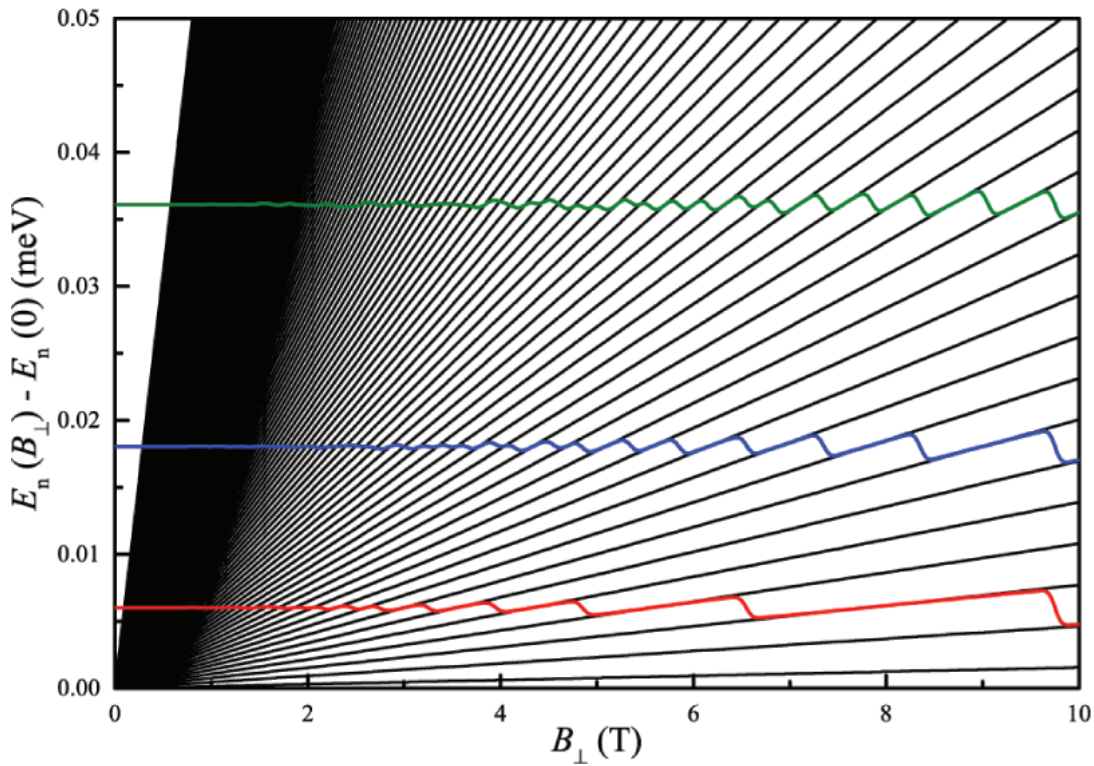}
		\caption{Fermi energy as a function of magnetic field, $B_{\perp}$, for $n_c = 1 \times 10^{16}$ m$^{-2}$ (red), $n_c = 3 \times 10^{16}$ m$^{-2}$ (blue), and $n_c = 6 \times 10^{16}$ m$^{-2}$ (green).}
	\label{Figure_EF}
\end{figure}

Fermi energy as a function of magnetic field is shown in Fig.~\ref{Figure_EF} for different electron concentrations $n_c$. For fixed Fermi energy, carrier concentration can be calculated by the formula, $n_{c}=\int_{0}^{E_{F}} D\left(E\right)dE$. Here, we see the magnetic field dependence of fixed Fermi energies for different carrier concentrations.

\begin{figure} [h]
    \centering
    \includegraphics[width=0.9\textwidth]{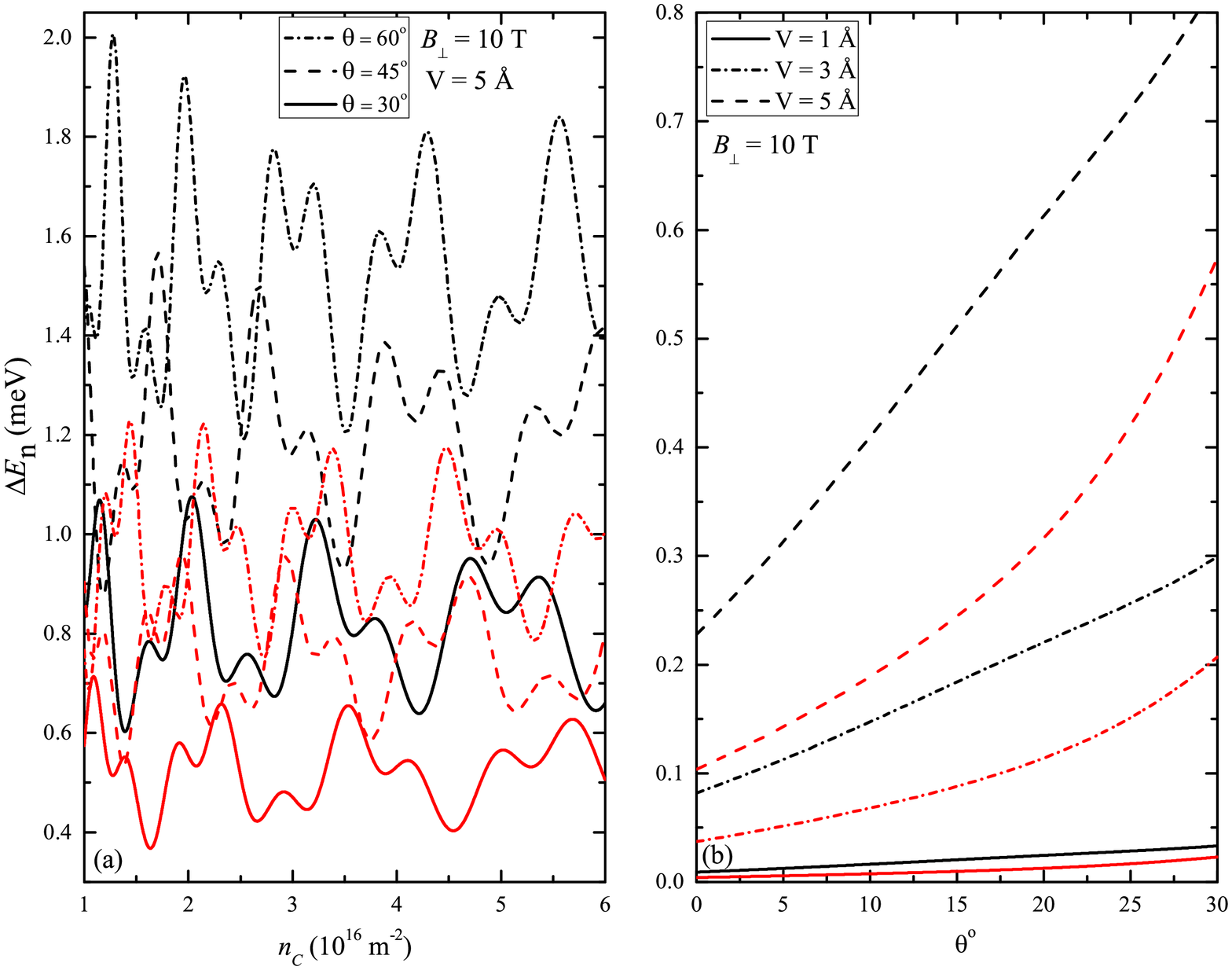}
    \caption{Tilted magnetic field contribution to the LL oscillations ($\Delta E_{n}$) calculated with respect to (a) carrier concentration $n_c$ for different $\theta$, and (b) $\theta$ for different $V$ at $n_{c}=1\times10^{16}$ m$^{-2}$. In all cases $\ell_{1}=\ell_{2}=100$ \AA{} and $\xi=0.5$. Black and red lines correspond to the electron and hole states, respectively.}
    \label{FIGURE5}
\end{figure}

To gain insight into the role of other model parameters on the LL spectrum, we analyze $\Delta E_{n}$ in more detail. 
In Fig.~\ref{FIGURE5}(a), $\Delta E_{n}$ is shown both for electrons and holes as a function of the carrier concentration $n_c$ calculated for different tilt angles $\theta$ at $B_{\perp}=10$ T. It can be seen that $\Delta E_{n}$ exhibits oscillations with $n_c$, and its amplitude increases for larger $\theta$. This behavior is attributed to the $\sin \theta$ factor in $\Delta E_{n}$ [Eq.~(\ref{DeltaE})]. The hole states turn out to be less affected by the magnetic field direction, which is due to the higher cyclotron mass. In Fig.~\ref{FIGURE5}(b), we show $\Delta E_{n}$ as a function of $\theta$ calculated for different $V$. According to Eq.~(\ref{DeltaE}), $\Delta E_{n} \sim V^2 (\sin \theta+\xi)^2$, meaning that the nonlinear effects in the spectrum of LLs increase both with $V$ and $\theta$. At small $\theta$, $\Delta E$ raises linearly, whereas at larger $\theta$, $\Delta E_n$ demonstrates a quadratic behavior.

\begin{figure} [h!]
	\centering
	\includegraphics[width=0.95\textwidth]{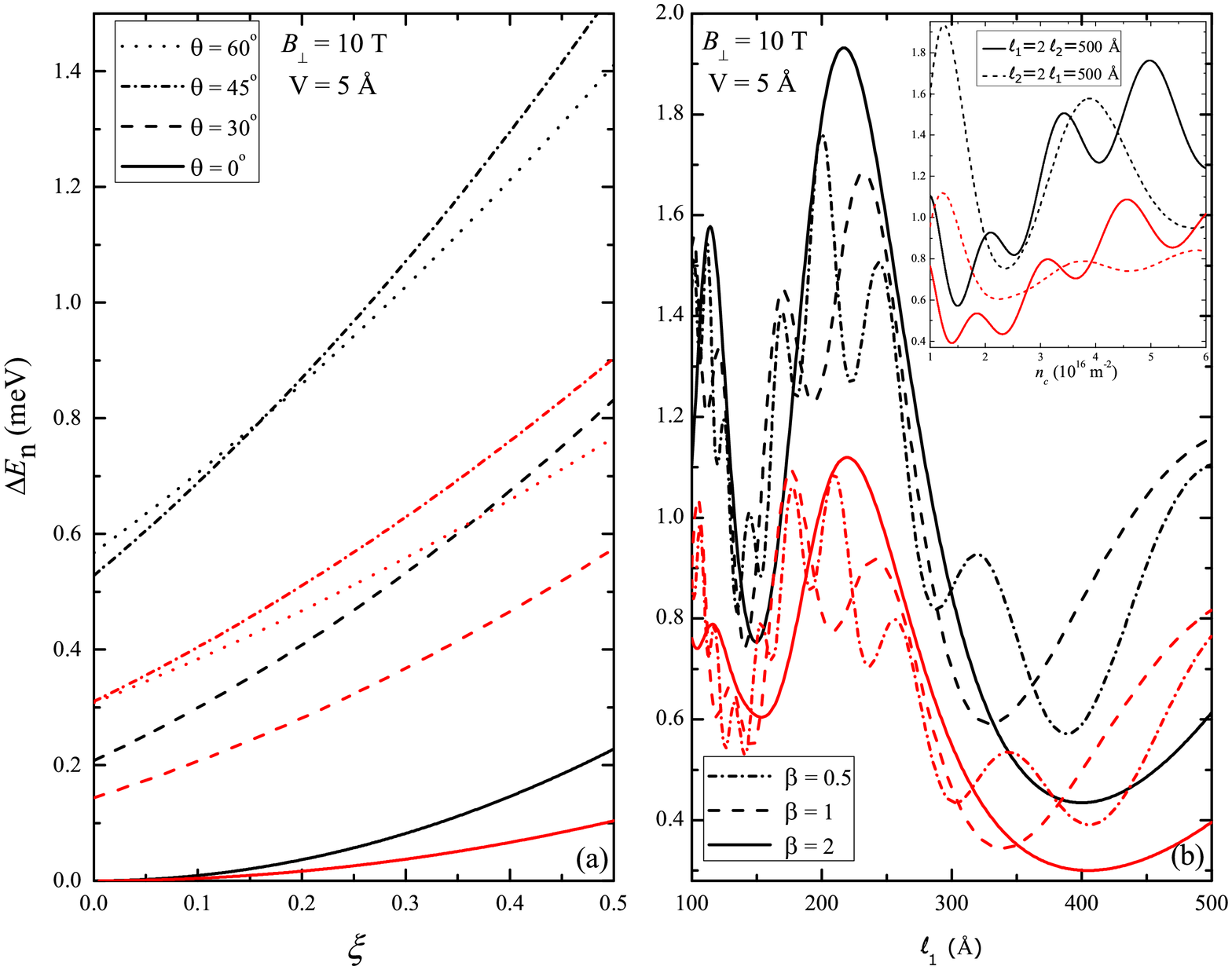}
		\caption{LL oscillations ($\Delta E_{n}$) calculated with respect to (a) $\xi=B_{\parallel}/B_{\perp}$ for different $\theta$, and (b) $\ell_1$ for different $\beta=\ell_{2}/\ell_{1}$ at $n_{c}=1 \times 10^{16} $ m$^{-2}$. The inset shows the dependence of $\Delta E_n$ on the carrier concentration $n_c$ for $\ell_{1}=2\ell_{2}=500$\AA{} and $\ell_{2}=2\ell_{1}=500$\AA{}. Black and red lines correspond to the electron and hole states, respectively.} 
	\label{FIGURE6}
\end{figure} 
    
In Fig.~\ref{FIGURE6}(a), we show the effect of a parallel magnetic field $B_{\parallel}$ by calculating the dependence of $\Delta E_n$ on the dimensionless parameter $\xi=B_{\parallel}/B_{\perp}$. One can see the expected from Eq.~(\ref{DeltaE}) $\Delta E_{n} \sim \xi^2$ behavior, suggesting that at large $\xi$ the in-plane field might play a role in the energy spectrum of corrugated MBP samples. The absolute effect of $B_{\parallel}$ is, however, not large and can hardly be detected experimentally under realistic field strengths and corrugation amplitudes. A similar effect of $B_{\parallel}$ on LLs was also reported previously in the context of bilayer graphene \cite{0953-8984-24-4-045501}. 
Fig.~\ref{FIGURE6}(b) shows the effect of the corrugation length $\ell_1$ as well as its anisotropy $\beta=\ell_2/\ell_1$ on $\Delta E_n$. In this case, $\Delta E_n$ exhibits a complicated oscillatory behavior. 
Keeping in mind anisotropic ripple formation typical to MBP\cite{doi:10.1021/acs.jpclett.5b00522}, we also examine $\Delta E_n$ as a function of $n_c$ for anisotropic corrugation patterns [see inset of Fig.~\ref{FIGURE6}(b)]. Depending on the corrugation direction, the behavior of LLs is significantly different. 
One can see, however, that $\Delta E_n$ remains weakly affected by a particular corrugation pattern as well as by the corrugation length.

\begin{figure} [h!]
	\centering
	\includegraphics[width=\textwidth]{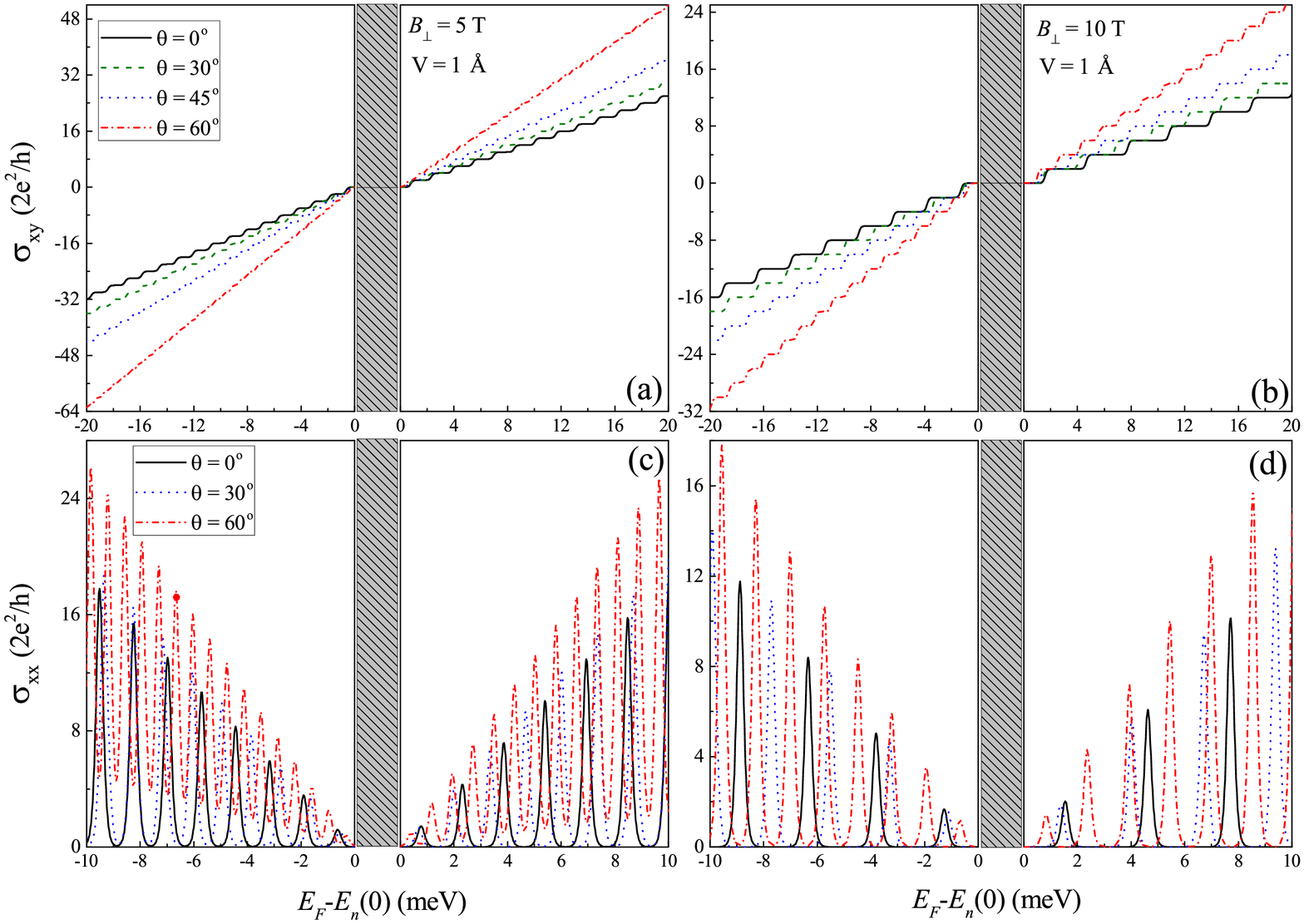}
		\caption{Hall conductivity ($\sigma_{xy}$) calculated for (a) $B_{\perp}=5$ T, (b) $B_{\perp}=10$ T, and longitudinal conductivity ($\sigma_{xx}$) calculated for (c) $B_{\perp}=5$ T, (d) $B_{\perp}=10$ T. Shaded area correspond to a gap in the energy spectrum.}
	\label{FIGURE7}
\end{figure} 

We now turn to the results of our magnetotransport calculations to see whether weak effects induced by the corrugation could be observed experimentally. The Hall ($\sigma_{xy}$) and longitudinal ($\sigma_{xx}$) conductivities are shown for different tilt angles in Fig.~\ref{FIGURE7}.
For Fermi energies in the gap region between the valence and conduction states, dc conductivity is obviously zero due to the absence of charge carriers. Beyond the gap region, $\sigma_{xy}$ exhibits distinct plateaus, arising from the discrete nature of the LL spectrum [Fig.~\ref{FIGURE7}(a)]. The Hall conductivity increases by $2e^2/h$ for each level forming the integer Hall plateaus indexed as $0,\pm2,\pm4,\pm6...$. It can be seen that $\sigma_{xy}$ increases with tilt angle, which is attributed to larger DOS caused by more dense LLs (cf. Fig.~\ref{FIGURE3}). For the same reason, $\sigma_{xy}$ becomes smaller in stronger fields [Fig.~\ref{FIGURE7}(b)]. At sufficiently small $V$, $\sigma_{xy} \sim (B_{\perp}\cos \theta)^{-1}$.
The longitudinal conductivity $\sigma_{xx}$ exhibits oscillatory behavior typical to the Shubnikov-de Haas (SdH) oscillations, as shown in Figs.~\ref{FIGURE7}(c) and (d). $\sigma_{xx}$ also increases with $\theta$ yet more slowly than $\sigma_{xy}$, because in this case $\sigma_{xx}\sim (B_{\perp}\cos \theta)^{-1/2}$ due to a factor $\overline{\ell}_{B}$ in the denominator of Eq.~(\ref{13}). 
We note that the effect of in-plane magnetic field ($B_{\parallel}$) and the corrugation lengths ($\ell_{1}$ and $\ell_{2}$) are negligible in the context of magnetotransport properties of MBP and, therefore, not presented here. The role of the corrugation amplitude is analyzed below.

\begin{figure} [h!]
	\centering
	\includegraphics[width=0.9\textwidth]{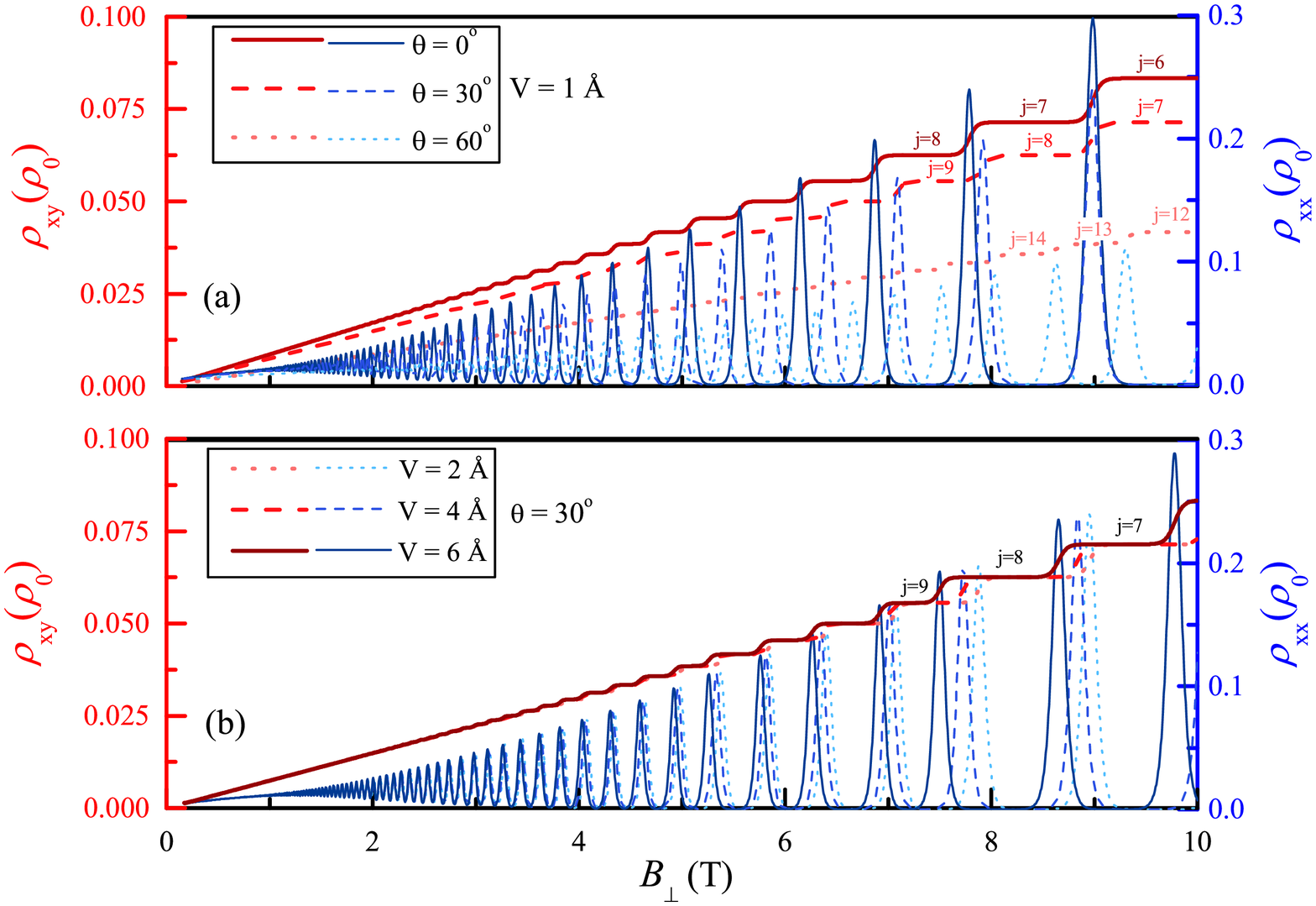}
		\caption{Hall resistivity ($\rho_{xy}$) and longitudinal resistivity ($\rho_{xx}$) versus magnetic field, $B_{\perp}$ for different tilt angles $\theta$. The case of electron doping with $n_c \approx 3 \times 10^{16}$ m$^{-2}$ is considered. $\rho_{0}=h/2e^2$ is the resistivity unit.}
	\label{FIGURE8}
\end{figure} 

In Fig.~\ref{FIGURE8}(a), the Hall $\rho_{xy}$ and longitudinal $\rho_{xx}$ resistivity are shown as a function of magnetic field calculated at different $\theta$ for the case of electron doping 
$n_c \approx 3 \times 10^{16}$ m$^{-2}$ and $V=1$ \AA{}. The behavior of $\rho_{xy}$ is closely related to $\sigma_{xy}$ shown in Fig.~\ref{FIGURE7}. As expected, $\rho_{xy}$ increases linearly with $B_{\perp}$, while larger $\theta$ correspond to effectively weaker fields. At large fields, $\rho_{xy}$ becomes quantized increasing by the unit of $\rho_{0}=h/2e^2$. 
For a given magnetic field, the Hall plateaus observed at different $\theta$ correspond to different filling factors $j$. The filling factors increase with $\theta$, meaning that the effect of the tilt angle is opposite to that of the Fermi energy, i.e., larger $\theta$ corresponds to shifting the position of the Hall plateaus toward smaller $E_F$ and vice versa.
Similar effect of tilted magnetic field on resistivities was reported for graphene\cite{doi:10.1063/1.4818605}. 
As can be seen from Fig.~\ref{FIGURE8}(a), the behavior of $\rho_{xx}$ is also similar to $\sigma_{xx}$, exhibiting pronounced SdH oscillations as well as a $\theta$-dependence of the peak amplitudes. One of the most interesting result is presented in Fig.~\ref{FIGURE8}(b), which shows the effect of the corrugation amplitudes $V$ on $\rho_{xy}$ and $\rho_{xx}$ calculated for a fixed $\theta$. Although the effect of $V$ is less pronounced compared to $\theta$, it becomes clearly seen at fields $B_{\perp}>5$ T. Larger $V$ shift the Hall plateaus in $\rho_{xy}$ as well as SdH oscillation peaks in $\rho_{xx}$ toward weaker magnetic fields. For sufficiently strong fields, the difference in corrugation amplitudes of a few \AA~results in a notable contraction of the $\rho(B_{\perp})$ spectrum along the field axis reaching 0.5 T at $B_{\perp} \sim 8$ T. Given that a realistic corrugation pattern would be represented by a superposition of different corrugation amplitudes, we expect a broadening of the SdH peaks increasing with $B_{\perp}$ under experimental conditions. Although such a behavior is typical to experimentally measured longitudinal and Hall resistivity in few-layer BP\cite{tayari2015,li2015quantum,li2016quantum,long2016,long2017}, it is
usually attributed to the Zeeman spin-splitting, well described by the standard Lifshitz-Kosevich formula for 2D resistivity. To reveal the effect of corrugations in the resistivity measurements, the broadening must increase with tilt angle, which is apparently not observed in known experiments on few-layer BP. The resolution of the available experimental spectra also does not allow us to observe nonlinear effects in the SdH oscillations. We note, however, that the existing magnetotransport measurements has been performed on a \emph{few-layer} BP, which should be significantly less affected by the structural corrugations compared to single-layer samples.

\section{Conclusions}
In summary, we studied the effects of tilted magnetic field and long-range structural corrugations on LLs and magnetotransport properties of MBP. We considered an analytical model and obtained first-order corrections to the LL energies induced by corrugations in the long-wavelength limit. The energies of LLs are found to be expectedly decreasing with the tilt angle due to the $\cos \theta$ factor in the modified cyclotron frequency. At sufficiently strong fields, however, the corrugation potential induces  nonlinear deviations in the dependence of LL energies on magnetic field. We find that these deviations are predominantly affected by the corrugation amplitude, whereas the corrugation length and its specific direction are less relevant.
We also examined the magnetotransport properties of MBP in the presence of corrugations under tilted magnetic field within the scheme of linear response theory. Overall, the tilt angle modifies the resistivity spectra considerably, effectively reducing the magnetic field strength. In the presence of long-range corrugations, we find that both Hall and longitudinal resistivity spectra display: (i) a shift toward weaker magnetic fields, and (ii) additional broadening of the SdH peaks increasing with magnetic field, not related to the Zeeman splitting. The obtained effects are noticeable even at moderate ($B<10$ T) fields, which allows us to expect that they might be observable experimentally for MBP samples deposited on sufficiently corrugated (e.g., SiO$_2$) substrates.

\begin{acknowledgments}
A. Mogulkoc would like to thank Professor B.~S. Kandemir for fruitful discussions. A.N.R. acknowledges support from the Ministry of Education and Science of the Russian Federation, Project No. 3.7372.2017/BP.
\end{acknowledgments}

\section*{References}

\bibliography{referans}
\appendix
\section{Derivation of energy eigenvalues in the presence of magnetic field}
In the symmetric gauge, energy eigenvalues of the Hamiltonian given by Eq.~(\ref{1}) can be expressed as

\begin{eqnarray}
E_{n\lambda}^{0}&=&u_{0}+\lambda \delta+(\bar{\eta}_x+\lambda \bar{\gamma}_x)\left[\frac{m^{\lambda}_{x} \hbar \omega_{\lambda}}{2}\langle(b_{x}^{\dagger}+b_{x})^{2}\rangle-\left(\frac{eB}{2}\right)^{2}\frac{\hbar}{2m^{\lambda}_{y}\omega_{\lambda}}\langle(b_{y}^{\dagger}-b_{y})^{2}\rangle\right]\notag \\
&+&(\bar{\eta}_y+\lambda \bar{\gamma}_y+\lambda \frac{\bar{\chi}^2}{2\delta})\left[\frac{m^{\lambda}_{y} \hbar \omega_{\lambda}}{2}\langle(b_{y}^{\dagger}+b_{y})^{2}\rangle-\left(\frac{eB}{2}\right)^{2}\frac{\hbar}{2m^{\lambda}_{x}\omega_{\lambda}}\langle(b_{x}^{\dagger}-b_{x})^{2}\rangle\right]\notag \\
&-&(\bar{\eta}_x+\lambda \bar{\gamma}_x)\left[ieB \sqrt{\frac{m^{\lambda}_{x} \hbar \omega_{\lambda}}{2}}\sqrt{\frac{\hbar}{2m^{\lambda}_{y}\omega_{\lambda}}}\langle(b_{x}^{\dagger}+b_{x}) (b_{y}^{\dagger}-b_{y})\rangle\right]\notag \\
&+&(\bar{\eta}_y+\lambda \bar{\gamma}_y+\lambda \frac{\bar{\chi}^2}{2\delta})\left[ieB \sqrt{\frac{m^{\lambda}_{y} \hbar \omega_{\lambda}}{2}}\sqrt{\frac{\hbar}{2m^{\lambda}_{x}\omega_{\lambda}}}\langle(b_{y}^{\dagger}+b_{y}) (b_{x}^{\dagger}-b_{x})\rangle\right],
\label{AA}
\end{eqnarray}%
where $\langle ... \rangle$ corresponds to expectation values between the oscillator states, $ |n_{x}n_{y}\rangle$. Creation and annihilation operators satisfy the commutation relation, $\left[b_{i},b_{j}^{\dagger}\right]=\delta_{ij}$ and they have eigenvalues $ b_{x(y)}|n_{x}n_{y}\rangle =\sqrt{n_{x(y)}}|n_{x-1(x)}n_{y(y-1)}\rangle$ and $ b_{x(y)}^{\dagger}|n_{x}n_{y}\rangle =\sqrt{n_{x(y)}+1}|n_{x+1(x)}n_{y(y+1)}\rangle$. Furthermore, number operators ($\hat{n}_{x}=b_{x}^{\dagger} b_{x}$ and $ \hat{n}_{y}=b_{y}^{\dagger} b_{y}$) have the following eigenvalues
\begin{eqnarray}
\hat{n}_{x}|n_{x}n_{y}\rangle&=& n_{x}|n_{x}n_{y}\rangle\notag \\
\hat{n}_{y}|n_{x}n_{y}\rangle&=& n_{y}|n_{x}n_{y}\rangle.
\label{AB}
\end{eqnarray}%
Here, $n_{x}$ and $n_{y}$ are positive integers, i.e., $n_{x(y)}=0,1,2...$. The last two terms in Eq.~(\ref{AA}) correspond to the angular momentum operator $L_{z}$, which satisfies the eigenvalue equation $ L_{z}|n_{x}n_{y}\rangle=\hbar\left(n_{y}-n_{x}\right)|n_{x}n_{y}\rangle$. Here, $n_{y}-n_{x}=m$ where $m$ is the magnetic quantum number. Using the assumption $n_{x}=n_{y}=n$, this term vanishes and Eq.(\ref{AA}) leads to Eq.~(\ref{6}). Further details can be found in Refs.~\onlinecite{0143-0807-28-1-002} and \onlinecite{messiah1961quantum}.
By the inclusion of tilted magnetic field and corrugation potential with the modified symmetric gauge, square of the momentum operators is given by Eq.~(\ref{Picorrug}). Similar method can be followed for the evaluation of the energy eigenvalues. Assuming $n_{x}=n_{y}=n$, non-zero elements of the energy eigenvalues can be written as,

\begin{eqnarray}
E_{n\lambda}^{0}&=&u_{0}+\lambda \delta+(\bar{\eta}_x+\lambda \bar{\gamma}_x)\left[\frac{m^{\lambda}_{x} \hbar \overline{\omega}_{\lambda}}{2}\langle(b_{x}^{\dagger}+b_{x})^{2}\rangle-\left(\frac{eB_{\perp}\cos \theta}{2}\right)^{2}\frac{\hbar}{2m^{\lambda}_{y}\overline{\omega}_{\lambda}}\langle(b_{y}^{\dagger}-b_{y})^{2}\rangle\right]\notag \\
&+&(\bar{\eta}_y+\lambda \bar{\gamma}_y+\lambda \frac{\bar{\chi}^2}{2\delta})\left[\frac{m^{\lambda}_{y} \hbar \overline{\omega}_{\lambda}}{2}\langle(b_{y}^{\dagger}+b_{y})^{2}\rangle-\left(\frac{eB_{\perp}\cos \theta}{2}\right)^{2}\frac{\hbar}{2m^{\lambda}_{x}\overline{\omega}_{\lambda}}\langle(b_{x}^{\dagger}-b_{x})^{2}\rangle\right]\notag \\
&+&(\bar{\eta}_x+\lambda \bar{\gamma}_x)\left[\frac{e^{2}B_{\perp}^{2}\left(\sin \theta+\xi \right)^{2}}{2}\frac{V^{2}}{4} \mathbb{G}_{n}\right]+(\bar{\eta}_y+\lambda \bar{\gamma}_y+\lambda \frac{\bar{\chi}^2}{2\delta})\left[\frac{e^{2}B_{\perp}^{2}\left(\sin \theta+\xi \right)^{2}}{2}\frac{V^{2}}{4} \mathbb{G}_{n}\right],
\label{AC}
\end{eqnarray}%
Here, first two terms result in $\bar{E}_{n\lambda}^{0}$ appearing in Eq.~(\ref{8}), which is a modified counterpart of Eq.~(\ref{7}) with $\cos \theta$ factor. The last term corresponds to the first-order correction to the energy eigenvalues given by Eq.~(\ref{DeltaE}). In turn, spatial correlation function $\mathbb{G}_{n}$ can be written using Eqs.~(\ref{9}) and (\ref{wf}) as

\begin{eqnarray}
\mathbb{G}_{n}&=&\left[\frac{1}{2^{n}n!\sqrt{\pi}}\sqrt{\frac{m^{\lambda}_{x} \overline{\omega}_{\lambda}} {\hbar}}\int_{-\infty}^{\infty}dx \cos^{2}(\textrm{K}x) \left|H_{n}\left(\sqrt{\frac{m^{\lambda}_{x} \overline{\omega}_{\lambda}} {\hbar}}x\right)\right|^{2}\right]\notag \\ &\times&\left[\frac{1}{2^{n}n!\sqrt{\pi}}\sqrt{\frac{m^{\lambda}_{y} \overline{\omega}_{\lambda}} {\hbar}}\int_{-\infty}^{\infty} dy \cos^{2}(\textrm{K}^{\prime}y)\left|H_{n}\left(\sqrt{\frac{m^{\lambda}_{y} \overline{\omega}_{\lambda}} {\hbar}}y\right)\right|^{2}\right].
\label{AD}
\end{eqnarray}%
Integrals in Eq.~(\ref{AD}) can be taken by using the identity \cite{gradshteyn2014table},

\begin{eqnarray}
\int_{0}^{\infty}d\alpha \exp \left(-\alpha^{2}\right)\left[H_{n} \left(\alpha\right)\right]^{2} \cos \left(\sqrt{2}\beta \alpha\right)&=& 2^{n-1}\sqrt{\pi}n! \exp \left(-\beta^{2}/2\right) L_{n}\left(\beta^{2}\right).
\label{AF}
\end{eqnarray}%
Here, $H_{n}(x)$ and $L_{n}(x)$ are the Hermite and the Laguerre polynomials, respectively. By using trigonometric identities [$\cos^{2}\theta=(1+\cos 2\theta)/2$] and Eq.~(\ref{AF}), one can recast spatial correlation function in Eq.~(\ref{AD}) as Eq.~(\ref{10}).

\section{Tight-binding description of magnetic field}

For the tight-binding calculations, we consider the model proposed in Ref.~\onlinecite{PhysRevB.89.201408}. The model consists of five hopping parameters between the $p_z$-like orbitals of phosphorus atoms. In the presence of external magnetic field the hopping parameters acquire a Peierls phase \cite{PhysRevB.51.4940}. The corresponding hopping between the atoms at $\mathbf{r}_1$ and $\mathbf{r}_2$ become

\begin{eqnarray}
t_{\mathbf{r}_1,\mathbf{r}_2}\rightarrow t_{\mathbf{r}_1,\mathbf{r}_2}\mathrm{exp}
\left({ie/\hbar \int_{\mathbf{r}_1}^{\mathbf{r}_2}\mathbf{A}\cdot d\mathbf{l}}\right),
\end{eqnarray}%
where $\mathbf{A}$ is the vector potential. For a homogeneous perpendicular magnetic field applied in the $z$ direction $B_{\perp}$, we choose the Landau gauge \textbf{A}=(0,$B_{\perp}x$,0). To preserve translation invariance of the system, the magnetic flux $\Phi$ through each unit cell
should be chosen as a rational multiply of the flux quantum $\Phi_0=e/h$ \cite{PhysRevB.51.4940}. In the case of MBP, the magnetic flux through each cell is
\begin{eqnarray}
\Phi=(e/h)\frac{B_{\perp}{a}_1{a}_2}{2},
\end{eqnarray}%
where $a_1$ and $a_2$ are lattice vectors in the $x$ and $y$ directions. In practice, it is convenient to consider a supercell composed of $q$ unit cells in the $x$-direction, such that $\Phi=\Phi_0/q$ \cite{PhysRevB.94.115118}. Therefore, low values of $B$ require large supercell meaning that the dimensionality of the tight-binding Hamiltonian increases as $B$ decreases. The LL are obtained by diagonalizing the Hamiltonian at the center ($\Gamma$ point) of the Brillouin zone.

\begin{figure} [h]
    \centering
    \includegraphics[width=0.7\textwidth]{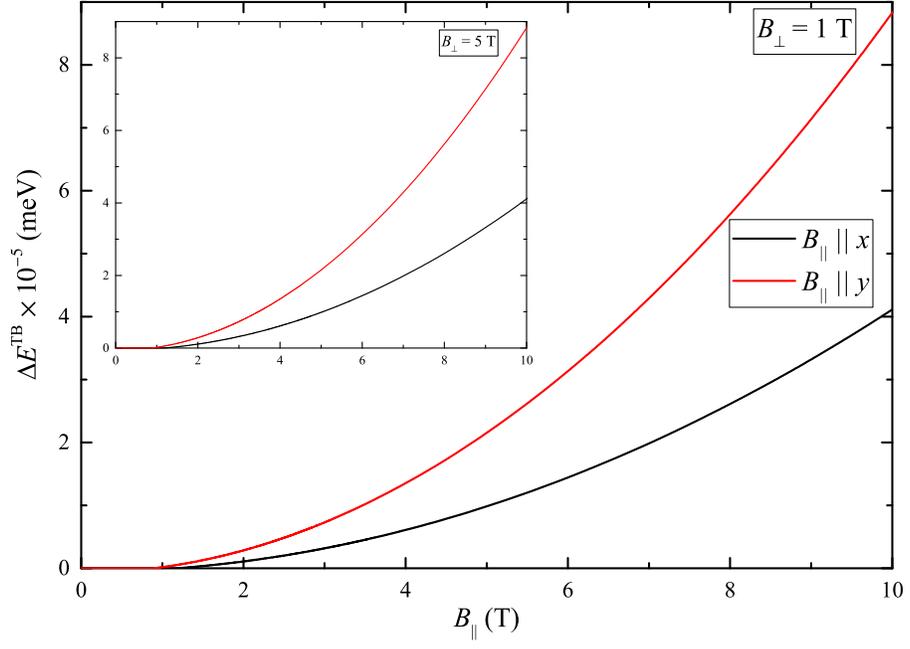}
    \caption{The contribution of in-plane field ($B_{\parallel}$) to the LL energies $\Delta E^{\mathrm{TB}}$ calculated for $B_{\perp}=1$ T and $B_{\perp}=5$ T (inset) for $n=0$.} 
    \label{FIGURE_B}
\end{figure}

For a perfectly planar atom-thick 2D material like graphene the in-plane magnetic flux through its structure is zero. This is generally not the case for 2D materials with finite thickness like bilayer graphene \cite{0953-8984-24-4-045501}. Here, for the puckered structure of MBP, the in-plane magnetic field induces a phase difference between the top and bottom sublayers of phosphorus atoms, which depends on the MBP thickness $d=2.1$ $\mbox{\AA}$. To study the effect of in-plane magnetic field on the electronic properties of MBP, we consider the in-plane field ($B_{\parallel}$) together with perpendicular magnetic field $(B_{\perp})$ using the vector potential \textbf{A}=(0,$B_{\perp}x$-$B_{\parallel}z$,0) and \textbf{A}=($B_{\parallel}z$,$B_{\perp}x$,0) for $x$- and $y$-directions of $B_{\parallel}$, respectively. Since there is no periodicity in the $z$-direction, the in-plane field does not produce any quantization due to the confinement of charge carriers within the $xy$ plane. However, it gives rise to field-dependent shifts of energy levels. In Fig.~\ref{FIGURE_B}, the contribution of in-plane field [$\Delta E^{\mathrm{TB}}=E(B_{\perp},B_{\parallel})-E(B_{\perp},0$)] to the LL energies is shown for moderate values of $B_{\parallel}$ applied along both $x$- and $y$-directions. Due to the relatively low buckling ($d=2.1$ $\mbox{\AA}$) and insignificant modification of tight-binding parameters (only $t_1$ and $t_3$ hoppings are primarily affected), the in-plane magnetic field has a negligible effect on LLs (order of $10^{-5}$ meV). This result is similar to bilayer graphene \cite{0953-8984-24-4-045501}, where noticeable changes appear only for $B_{\parallel}>50$ T. Here, however, one can see the anisotropy of contribution due to the direction-dependent effective masses in MBP.    
We also analyzed the $n$ dependence of $\Delta E^{\mathrm{TB}}$ (not shown here) and conclude that the effect of $B_{\parallel}$ is almost uniform and the $n$ dependence can be considered as negligible.

\end{document}